\documentclass[epj]{svjour}
\usepackage{graphicx}
\newcommand{\infnpd}{1}
\newcommand{\edinburgh}{3}
\newcommand{\unina}{4}
\newcommand{\lngs}{5}
\newcommand{\dresden}{6}
\newcommand{\infnmi}{8}
\newcommand{\infnge}{7}
\newcommand{\dippd}{9}
\newcommand{\atomki}{10}
\newcommand{\infnto}{11}
\newcommand{\unipd}{12}
\newcommand{\infnLNL}{2}
\newcommand{\bochum}{13}
\newcommand{\teramo}{14}
\newcommand{\caserta}{15}

\begin{document}

\title{Preparation and characterisation of isotopically enriched Ta$_2$O$_5$ targets for nuclear astrophysics studies}

\author{
	A.\,Caciolli \inst{\infnpd,\infnLNL}\thanks{e-mail: caciolli@pd.infn.it} \and
	D.\,A.\,Scott \inst{\edinburgh} \and
	A.\,Di Leva \inst{\unina} \and
	A.\,Formicola \inst{\lngs}  \and
	M.\,Aliotta \inst{\edinburgh} \and
	M.\,Anders \inst{\dresden} \and
	A.\,Bellini \inst{\infnge} \and
	D.\,Bemmerer \inst{\dresden} \and
	C.\,Broggini \inst{\infnpd} \and
	M.\,Campeggio \inst{\infnmi} \and
	P.\,Corvisiero \inst{\infnge} \and
	R.\,Depalo \inst{\dippd} \and
	Z.\,Elekes \inst{\dresden} \and
	Zs.\,F\"ul\"op \inst{\atomki} \and
	G.\,Gervino \inst{\infnto} \and
	A.\,Guglielmetti \inst{\infnmi} \and
	C.\,Gustavino \inst{\lngs} \and
	Gy.\,Gy\"urky \inst{\atomki} \and
	G.\,Imbriani \inst{\unina} \and
	M.\,Junker \inst{\lngs} \and
	M.\,Marta \inst{\dresden}\thanks{\emph{Present Adress:} GSI Helmholtzzentrum f\"ur Schwerionenforschung GmbH, 64291 Darmstadt, Germany} \and
	R.\,Menegazzo \inst{\infnpd} \and
	E.\,Napolitani \inst{\unipd} \and
	P.\,Prati \inst{\infnge} \and
	V.\,Rigato \inst{\infnLNL} \and
	V.\,Roca \inst{\unina} \and
	C.\,Rolfs \inst{\bochum} \and
	C.\,Rossi Alvarez \inst{\infnpd} \and
	E.\,Somorjai \inst{\atomki} \and
	C.\,Salvo \inst{\lngs,\infnge} \and
	O.\,Straniero \inst{\teramo} \and
	F.\,Strieder \inst{\bochum} \and
	T.\,Sz\"ucs \inst{\atomki} \and 
	F.\,Terrasi \inst{\caserta} \and
	H.P.\,Trautvetter \inst{\bochum} \and
	D.\,Trezzi \inst{\infnmi} \\
	(LUNA collaboration)
	}
%
%
\institute{
	INFN, Sezione di Padova, via Marzolo 8, 35131 Padova, Italy 
	\and 
	INFN, Laboratori Nazionali di Legnaro, Legnaro, Italy 
	\and
	SUPA, School of Physics and Astronomy, University of Edinburgh, Edinburgh, UK  
	\and
	Dipartimento di Scienze Fisiche, Universit\`a di Napoli ``Federico II'', and INFN, Sezione di Napoli, Napoli, Italy 
	\and
	INFN, Laboratori Nazionali del Gran Sasso, Assergi, Italy 
	\and
	Helmholtz-Zentrum Dresden-Rossendorf, Dresden, Germany 
	\and
	Dipartimento di Fisica, Universit\`a di Genova, and INFN, Genova, Italy 
	\and
	Universit\`a degli Studi di Milano and INFN, Sezione di Milano, Milano, Italy 
	\and 
	Dipartimento di Fisica e Astronomia, Universit\`a di Padova, and INFN, Sezione di Padova, Via Marzolo 8, Padova, Italy 
	\and
	Institute of Nuclear Research (ATOMKI), Debrecen, Hungary 
	\and
	Dipartimento di Fisica Sperimentale, Universit\`a degli Studi di Torino, and INFN, Sezione di Torino, Torino, Italy 
	\and
	MATIS-IMM-CNR at Dipartimento di Fisica e Astronomia, Universit\`a di Padova, Via Marzolo 8, Padova, Italy 
	\and
	Institut f\"ur Experimentalphysik III, Ruhr-Universit\"at Bochum, Bochum, Germany 
	\and
	Osservatorio Astronomico di Collurania, Teramo, and INFN, Sezione di Napoli, Napoli, Italy 
	\and
	Seconda Universit\`a di Napoli, Caserta, and INFN, Sezione di Napoli, Napoli, Italy 
	}

\date{\today}
%
\abstract{
The direct measurement of reaction cross sections at astrophysical energies often requires the use of solid targets of known thickness, isotopic composition, and stoichiometry that are able to withstand high beam currents for extended periods of time. Here, we report on the production and characterisation of isotopically enriched Ta$_2$O$_5$ targets for the study of proton-induced reactions at the Laboratory for Underground Nuclear Astrophysics facility of the Laboratori Nazionali del Gran Sasso. The targets were prepared by anodisation of tantalum backings in enriched water (up to 66\% in $^{17}$O and up to 96\% in $^{18}$O). Special care was devoted to minimising the presence of any contaminants that could induce unwanted background reactions with the beam in the energy region of astrophysical interest. Results from target characterisation measurements are reported, and the conclusions for proton capture measurements with these targets are drawn.
} 

\authorrunning{A.\,Caciolli {\it et al.} (LUNA collab.)}
\titlerunning{Enriched Ta$_2$O$_5$ targets}

\maketitle

\section{Introduction}
\label{intro}

Stars spend most of their lives converting hydrogen into helium through a sequence of reactions known as the pp-chain  (for stars of mass M$\leq$1.5 M$_\odot$) or the CNO cycles (mainly for stars of mass M$>$1.5 M$_\odot$)  \cite{RolfsBook,Iliadis2007}.
An extensive experimental programme to study key reactions for hydrogen burning in stars has been carried out over the last two decades at the Laboratory for Underground Nuclear Astrophysics (LUNA) facility in Italy \cite{Costantini2009,Broggini2010}. 

Recently, the LUNA collaboration has undertaken the study of proton-induced reactions on $^{17}$O and $^{18}$O,  important nucleosynthesis processes in several stellar sites, including red giants, asymptotic giant branch (AGB) stars, massive stars, and classical novae \cite{pal10,Iliadis2002}.   
In particular, the ratio between the rates of the $^{17}$O(p,$\alpha$)$^{14}$N reaction ($Q$ = 1191.8 keV) and the $^{17}$O(p,$\gamma$)$^{18}$F reaction ($Q$ = 5606.5 keV) affects the galactic abundance of $^{17}$O, the stellar production of the radioactive $^{18}$F nuclide, and the predicted oxygen isotopic ratios in pre-solar grains \cite{pal10}. For the study of such reactions, solid targets enriched in $^{17}$O and $^{18}$O isotopes have been made. 

In the experimental investigation of reaction cross sections at the sub-Coulomb energies of astrophysical interest, solid targets have to meet a number of specific requirements \cite{seuthe1987}. These include: i) suitable and stable thickness, so as to sustain high beam currents (typically several hundreds of $\mu$A) over extended periods of time (from several hours up to a few days); 
ii) known and constant stoichiometry, so as to allow for accurate beam energy-loss calculations; and iii) known and possibly enriched isotopic composition, so as to allow for measurable yields. Each of these features should be measured to a high degree of accuracy to allow for the correct interpretation of reaction yields and thus of cross-section data. Finally, the target properties should be carefully monitored since changes in thickness, stoichiometry and/or composition under beam irradiation directly affect the measured absolute cross sections and may lead to significant systematic errors \cite{RolfsBook,Iliadis2007}.

The results of the $^{17}$O(p,$\gamma$)$^{18}$F measurements carried out at LUNA will be reported in a forthcoming paper \cite{scott2012}. Here, we present the procedure used to prepare solid Ta$_2$O$_5$ targets with an isotopic enrichment in $^{17}$O or $^{18}$O up to 66\% and 96\%, respectively, together with results on target characterisation using: i) Nuclear Resonant Reaction Analysis (NRRA); ii) Rutherford Backscattering (RBS); and iii) Secondary Ion Mass Spectrometry (SIMS) techniques. 

The implications of these results for the $^{17}$O(p,$\gamma$)$^{18}$F cross-section measurements are also briefly discussed.

\section{Experimental procedure}
\subsection{Target preparation}

Solid $^{17}$O- and $^{18}$O-enriched targets were prepared by anodic oxidation of tantalum backings in isotopically enriched water, a technique known to produce targets with highly uniform stoichiometry and homogeneous 
thick\-nesses \cite{Philips}. Tantalum disks (40 mm diameter) were obtained from 0.3  mm thick Ta sheets (99.9\% nominal purity) provided by Goodfellow Ltd.. The disks were initially polished by dipping the specimens in a solution of H$_2$SO$_4$, HNO$_3$ and HF (5:2:2). After the cleaning process, several targets show\-ed traces of light elements such as $^{11}$B,$^{12}$C, and $^{19}$F during irradiation with proton beam at LUNA.  In particular, $^{19}$F gives rise to an intense $\gamma$-ray background in the region of interest for the $^{17}$O(p,$\gamma$)$^{18}$F and $^{18}$O(p,$\gamma$)$^{19}$F reactions. Thus, a renewed effort was devoted to reduce the fluorine contamination from the tantalum backings. The new adopted procedure consisted in etching the tantalum backings in a bath of  20\% citric acid solution in water for 1h at 90$^\circ$C, before proceeding with the anodising process described below. The fluorine contamination in the $\gamma$ spectra was then observed to be reduced by a factor of 3.

\begin{figure}[!htb]
\begin{center}
\includegraphics[width=\columnwidth]{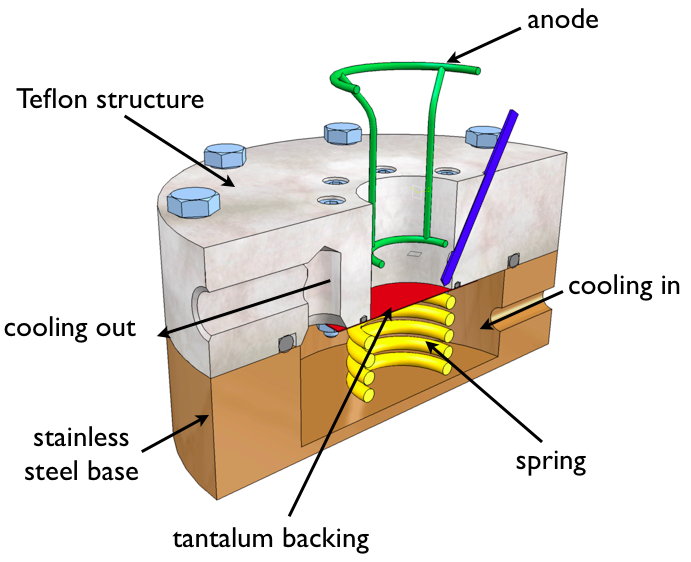}
\end{center}
\caption{(Colour online) Sketch of the anodisation device, consisting of a teflon annulus mounted on a stainless steel base. A tantalum disk placed at the interface between the teflon and the stainless steel base acted as the cathode while a gold-nickel ``basket" inserted in the hole acted as the anode. The cavity in the teflon defines the volume where the electrolytic solution was placed. A refrigerating fluid kept at constant temperature was circulated inside the device
to avoid evaporation of the solution during the anodisation process.}
\label{fig:schema1}
\end{figure}

The anodising apparatus consisted of a cylindrical annulus of teflon (100 mm and 25 mm outer and inner diameter, respectively, 25 mm height) fixed on a cylindrical base of stainless steel, as shown in fig. \ref{fig:schema1}. 
The teflon was chosen for its thermal stability and good electrical insulation.  
The cavity in the teflon defines the effective volume of
the anodising cell where the electrolytic solution is placed. A
tantalum disk placed at the bottom of the teflon cylinder acted as the
cathode and was maintained at ground potential through contact with a metallic spring.
The electrolyte consisted of a 0.1 mol potassium iodide (KI) salt solution with $^{17,18}$O enriched water (Cambridge Isotope Laboratories Inc.).
The KI electrolyte was chosen because
the anion contains no oxygen that could affect the final stoichiometric
ratio of the anodised Ta target and its salt crystallises without
water hydration. A basket (90\% gold and 10\% nickel) of 2 cm diameter acted as the anode (fig. \ref{fig:schema1}). Its vertical position could be suitably adjusted, in co-axial symmetry to the cathode, in order to reach the electrolyte surface. This served to minimise bubble production inside the electrolyte and thus to allow for more homogeneous deposition growths.

A stabilised DC power supply connected to the anode was monitored through a PC running LabView software controlling the ramp up and down of the DC power supply in order to keep the current inside the electrolyte solution constant.
Unwanted loss of the enriched water by evaporation, due to the heat produced during anodisation, was minimised by cooling the anodisation device so as to maintain the deposition process at a constant temperature to within $\pm$ 0.1 $^\circ$C.

\subsection{Oxide film growth}

Tests on oxide film growth were carried out as a function of the electrolyte solution temperature in the range $T= 17-27 ~^\circ$C, in steps of $2 ~^\circ$C. The observed final deposition thickness $d$ and rate of growth depend sensitively on temperature, according to a phenomenological  relationship \cite{Vermy1953,Mott1948}:
\begin{equation}\label{eq1}
\frac{I}{d} = a-b \ln t
\end{equation}
where $I$ is the current flowing through the electrolyte, $t$ the time of growth, and $a$ and $b$ temperature-dependent parameters. The growth rate is initially very high, but rapidly decreases to a maximum achievable thickness for a given temperature.  Depositions with best growth rates and homogeneity were obtained at 25$^\circ$C, where both current and voltage remained stable, i.e. with no spikes.
Successful oxide growths were achieved in the 40$-$220 V voltage range, with an estimated growth rate of  $\sim$32 {\rm \AA}/V, at a constant current density of 2mA/cm$^2$,  in qualitative and quantitative agreement with \cite{Seah1988} and \cite{Philips}, respectively.
Targets of various thicknesses ($1200-5800~{\rm \AA}$) and isotopic enrichment were produced using two solutions: solution 1 contained $^{17}$O (66\%) + $^{18}$O (4\%) + $^{16}$O (30\%); and solution 2 contained $^{18}$O (96\%) + $^{16}$O (4\%)  (targets enriched in $^{18}$O have not been used at LUNA yet and are not further discussed in this paper).

It should be noted, however, that for the present measurements \cite{scott2012} of the $^{17}$O(p,$\gamma$)$^{18}$F reaction only $^{17}$O-enriched targets of thickness less than 2000 {\rm \AA} were used.
At beam energies\footnote{All energies are quoted in the laboratory frame unless otherwise stated.} of astrophysical interest ($E_{\rm lab}=180-400~$keV) this thickness corresponds to beam energy losses $\Delta E <  35$ keV, {\em i.e.} a value appropriate for the measurement of both non-resonant and resonant contributions from the $^{17}$O(p,$\gamma$)$^{18}$F reaction. In this latter case, the resonant yield at $E_{\rm lab}=193~$keV can be safely measured without contributions from the more intense resonance at $E_{\rm lab}=151~{\rm keV}$ in the $^{18}$O(p,$\gamma$)$^{19}$F reaction arising from small $^{18}$O contaminations in the targets (see section \ref{sect:nuc}). 

\section{Target thickness measurements}
\label{sect:nuc}

Accurate determinations of the target thicknesses were obtained by means of Nuclear Resonant Reaction Analysis (NRRA), using a thick-target yield approach \cite{RolfsBook,Iliadis2007} during the actual proton-induced reaction studies at LUNA. Thick-target yield measurements are typically used to extract information on the location and strength of narrow resonances. However, for known resonances, the method can be used to provide an accurate measurement of the target thickness and provides a way of assessing the degree of deterioration under heavy beam bombardment \cite{seuthe1987}.

Target thickness measurements were performed concurrently with the study of proton-induced reactions at LUNA using the well-known, isolated, and narrow resonance of the $^{18}$O(p,$\gamma$)$^{19}$F reaction (Q=7994.8 keV)  at $E_{\rm lab}=151~{\rm keV}$. The resonance has a strength  $\omega \gamma$ = 1.0$\pm$0.1 meV and a total width $\Gamma <$ 300 eV \cite{Wiescher1980}.
Targets enriched in $^{17}$O were doped in $^{18}$O (at the level of 4\%) to allow for measurements of the $^{18}$O(p,$\gamma$)$^{19}$F resonance in a reasonable time. 

\begin{figure}[!h]
\begin{center}
\includegraphics[width=\columnwidth]{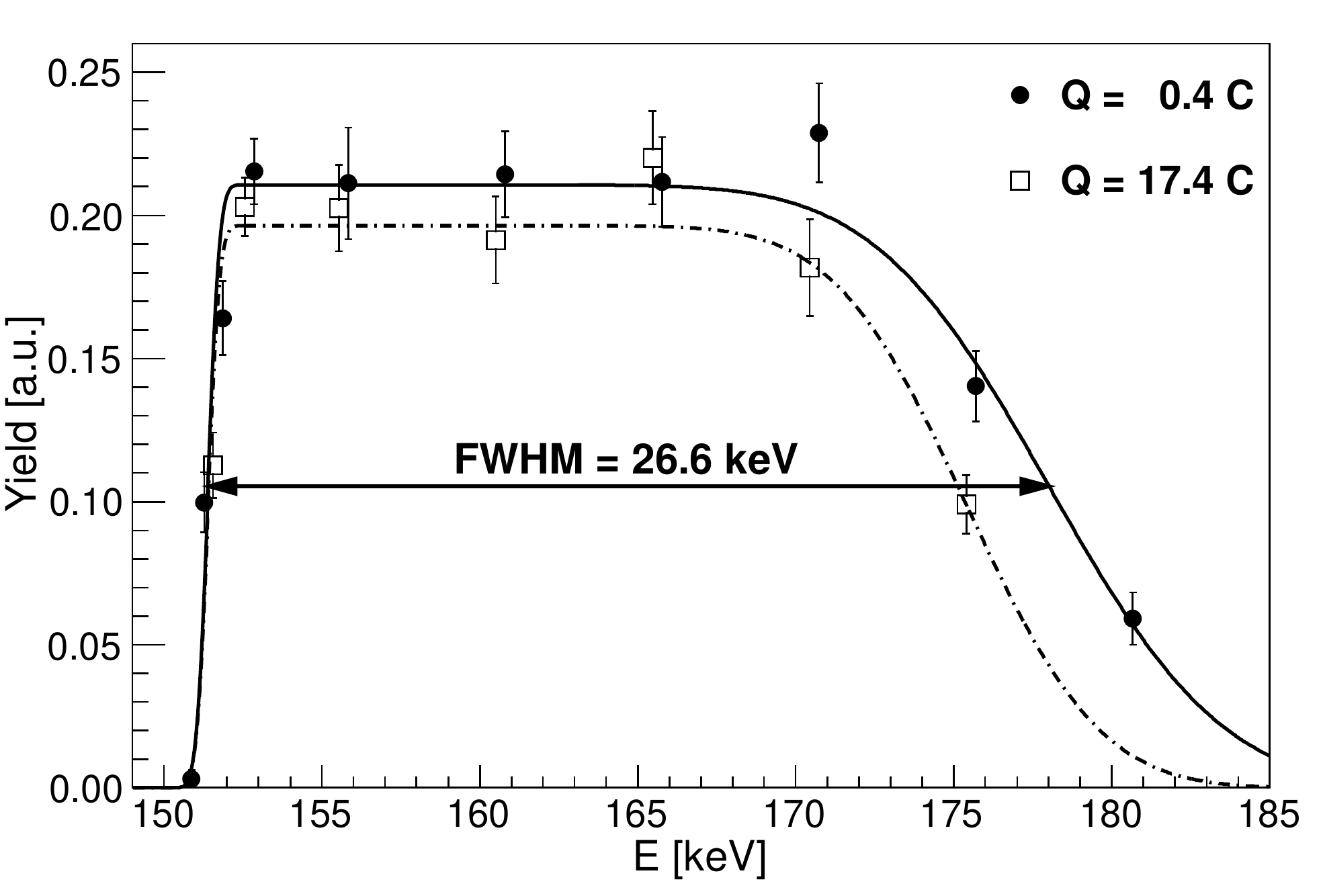}
\end{center}
\caption{Thick-target yield profile for the 151 keV resonance in $^{18}$O(p,$\gamma$)$^{19}$F as a function of beam energy ( errors are statistical only). A fit to the profile (here for target 24) leads to a target thickness $\Delta E = 26.6 \pm 0.7$ keV. The dashed line is a fit to the thick-target yield profile obtained after a total accumulated charge on target of 17.4 C and displays  a reduction in thickness. Each fit is affected by a 4\% error and thus the profile heights are broadly in agreement.}
\label{fig:thinscan}
\end{figure}

The targets were bombarded by a proton beam (typical intensity $I \sim 200\mu$A,  1 cm$^2$ spot size) in the energy range $E_{\rm lab}$ = 150-180 keV. At each beam energy, the $^{18}$O(p,$\gamma$)$^{19}$F resonant yield was determined by measuring the resonance de-excitation $\gamma$ rays at energies $E_\gamma = 3910$ and 4235 keV with a HPGe detector, 115\% relative efficiency, placed at 1.5 cm from the target. Both the target and the detector were at and angle of 55$^\circ$ with respect to the beam direction. A typical thick-target yield profile is shown in fig. \ref{fig:thinscan}. 

The yield curve displays a sharp rise at low-energy, whose slope arises from a combination of beam energy spread and resonance width. The smoother drop at high-energy is mostly due to the beam straggling in the target.
Since both the beam energy spread (typically $\Delta_{\rm beam} \sim $ 100 eV at LUNA \cite{Formicola2003}) and the beam straggling ($\Delta_{\rm stragg} \sim$ 2-4 keV \cite{SRIM} depending on  target) are much less than the observed full width at half maximum (FWHM), the latter can be taken as a measure of the target thickness $\Delta$E \cite{Iliadis2007,Fowler}.
Values of FWHM, {\em i.e.} $\Delta$E, were obtained through a fit of the yield curves using two error functions, $\mbox{erf}$:
\begin{equation}\label{eqErf}
F(E) = a0\left[\left({\rm erf} \left(\frac{E-a1}{a2}\right) + 1\right)  \left({\rm erf}\left( \frac{a3-E}{a4}\right) + 1\right)\right]
\end{equation}
where the target thickness is deduced from the difference between the position of the rising ($a1$) and the falling  ($a3$) edges. The $a2$ and $a4$ parameters reflect the beam energy spread and straggling at the entrance and  exit of the target, respectively.

\begin{figure}[!h]
\begin{center}
\includegraphics[width=.48\textwidth]{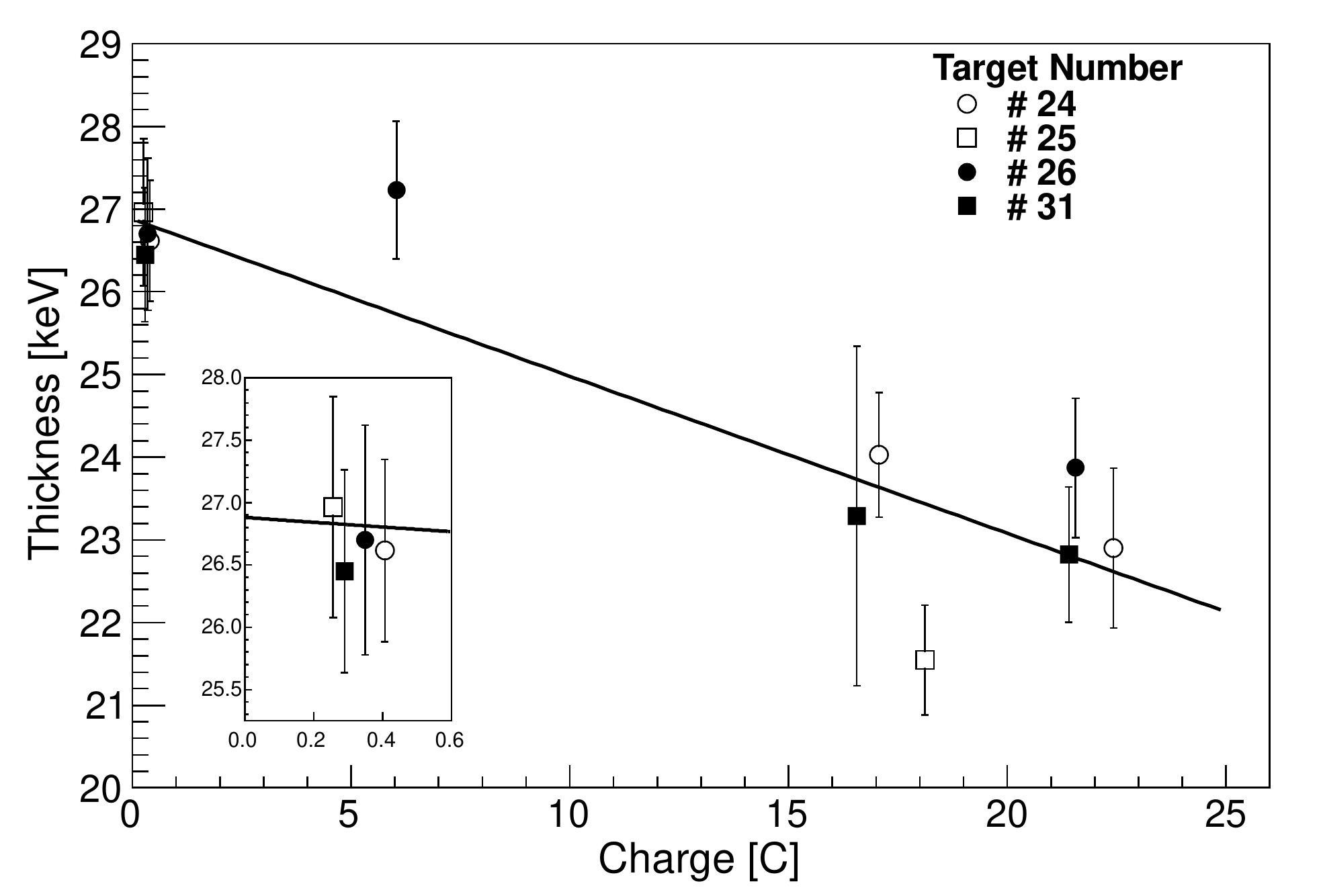}
\caption{Target thickness as a function of accumulated charge (here targets from the same batch). The solid line is a linear best fit to the data.} 
\label{fig:ThickVsCharge}
\end{center}
\end{figure}

To investigate the extent of target deterioration under heavy beam bombardment, measurements of the thick-target yield profiles were repeated as a function of the accumulated charge on target (fig. \ref{fig:ThickVsCharge}). 

The target thickness was found to decrease linearly with increasing charge for all targets investigated. Typically, a reduction rate of $\sim 0.2$ keV/C was observed, with a standard deviation increasing from 3.0\%  to 4.4\% with increasing charge, up to 25 C, as determined by a linear best fit to the experimental data. The accuracy of the fitting procedure was limited by the statistical precision being constrained by time considerations and/or  the comparatively small amounts of $^{18}$O in  the targets.

The thickness reduction is most likely due to sputtering induced by the proton beam. Additional factors ({\em e.g.} beam focalisation properties, heat dissipation effects) may contribute in different ways to the unique deterioration pattern of individual targets, and thus to the relatively large scatter of points around the best-fit line. 

For measurements of the proton-induced reaction cross-sections, maximum accumulated charges were limited to 25 C, corresponding to at most a 20\% degradation in target thickness.

\section{Stoichiometry measurements}
\label{sect:rbs}

For nuclear astrophysics applications, a key parameter of compound solid targets is the stoichiometric ratio of their constituents. The target stoichiometry determines the target stopping power and therefore the effective beam energy associated to the measured reaction cross section of astrophysical interest \cite{Iliadis2007}.  Here, the stoichiometry was obtained using the Rutherford Backscattering (RBS) technique \cite{Chu1978}.
The measurements were carried out at the AN2000 accelerator of Laboratori Nazionali di Legnaro (LNL) using a $^4$He beam (1 mm$^2$ spot size) at $E_{\rm lab}$ = 1.8 and 2.0 MeV.   
Scattered $\alpha$-particles were detected with a $25~{\rm mm}^2$ silicon detector placed at $\theta = 160^\circ$ with respect to the beam axis. The beam current (typically 90 nA) was monitored  by a Faraday cup. 

\begin{figure}[!htb]
\begin{center}
\includegraphics[width=\columnwidth]{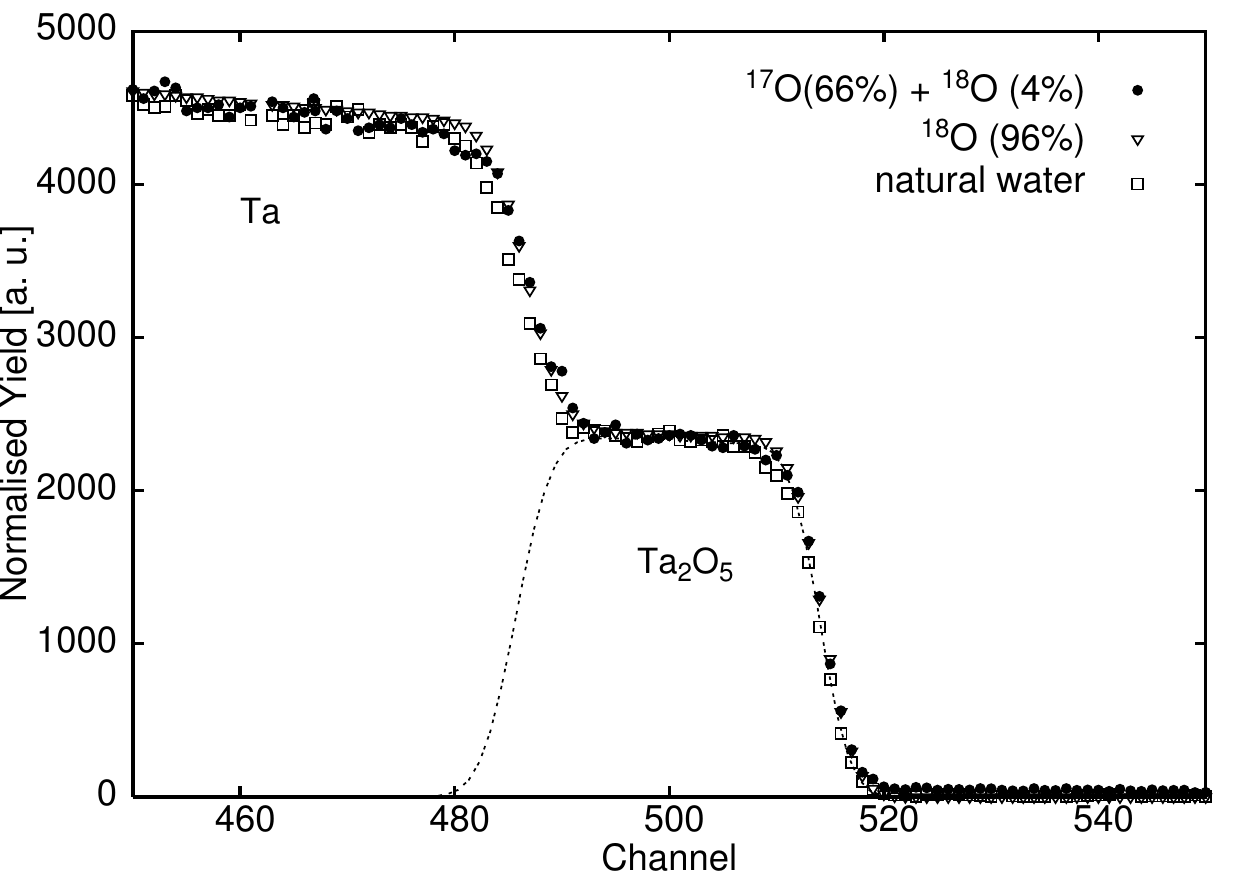}
\end{center}
\caption{RBS spectra for three targets prepared with different solutions. The Ta/O ratio is independent of the solution used as shown by the same relative height of the step and the bulk yields for each curve. The simulated tantalum oxide layer thickness  (dashed line) is in excellent agreement with the experimental data (see text for details).}
\label{fig:RBS1}
\end{figure}

A typical RBS spectrum is shown in fig. \ref{fig:RBS1} for three fresh targets ({\em i.e.} not previously exposed to the proton beam) prepared under the same growth conditions using three different electrolyte solutions (pure natural water;  water enriched in $^{17}$O; and water enriched in $^{18}$O). 

Simulations performed with SIMNRA 6.0 \cite{SIMNRA} (dashed line in the figure), taking into account both pile-up and dead time corrections and assuming Bragg type stopping power, give the overall Ta and O content in the oxide layer as 2.1$\times 10^{17}$ Ta/cm$^2$  and $5.4 \times 10^{17}$ O/cm$^2$, respectively, leading to a Ta/O ratio of $0.39\pm 0.02$\footnote{The uncertainty was determined from the averages of RBS 
results on several targets.}, in very good agreement with an expected value 0.4 \cite{Philips,Philips_2}. It is worth mentioning that a 5\% uncertainty on the stoichiometry leads to a 3\% systematic error on the effective stopping power and, thus, on the resonance strength. The stoichiometric ratio was observed to be the same for each of the electrolyte solutions within the uncertainties of the RBS technique. 

\begin{figure}[!htb]
\begin{center}
\includegraphics[width=\columnwidth]{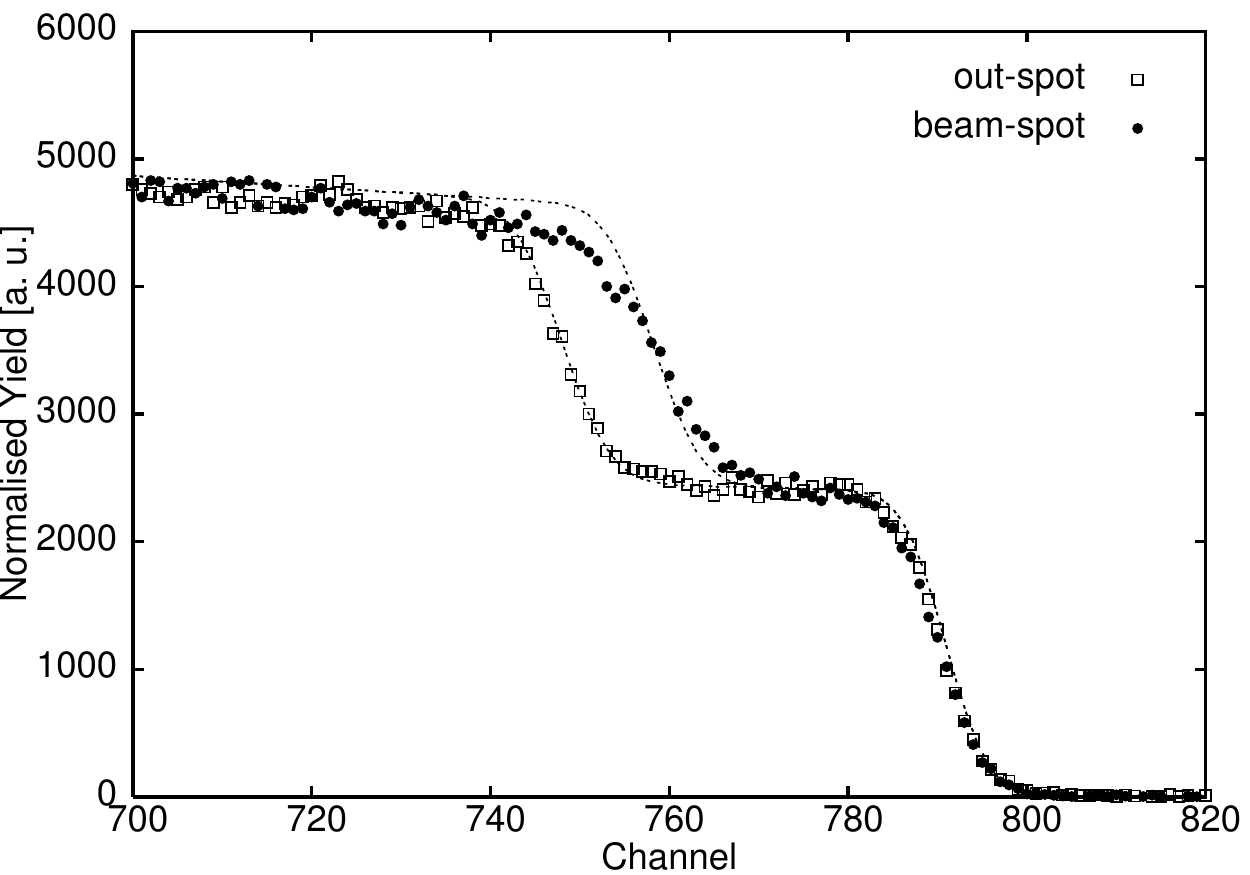}
\end{center}
\caption{Comparison of RBS yields for measurements in different regions (beam-spot: exposed to proton beam bombardment; out-spot: not exposed) of the same target. The beam-spot curve shows a reduced width of the Ta$_2$O$_5$ layer (step-wise structure), thus indicating a target deterioration, in qualitative agreement with thick-target yield measurements (fig. \ref{fig:thinscan}). The stoichiometric Ta/O ratio was not affected by the proton beam bombardment.}
\label{fig:RBS2}
\end{figure}

In order to study possible changes in the target stoichiometry under proton beam bombardment, RBS measurements were carried out  both in the target central area exposed to the LUNA proton beam (``beam-spot'') and on peripheral regions of the target where the beam intensity was negligible (``out-spot''). A typical example is shown in fig. \ref{fig:RBS2} for a target (target 24) exposed to a total accumulated charge of 23 C.
Although a partial deterioration in the target thickness is evident (in qualitative agreement with thick-target yield measurements), the stoichiometric ratio was not affected by the beam bombardment as indicated by the same relative step-to-maximum yields in both beam-spot and out-spot curves.  
From the out-spot curve, a Ta$_2$O$_5$ layer of 7.6$\times 10^{17}$ atoms/cm$^2$ (corresponding to $\Delta E$ = 27 keV at 151 keV proton beam, for a target tilted at 55$^o$) was determined, in excellent agreement with the thick-target yield results (see fig. \ref{fig:thinscan}).

\section{Isotopic enrichment measurements}

Another critical parameter for the proton-induced reaction studies at LUNA, was the oxygen isotopic  abundance. 
Depth profiles of isotopic abundances of $^{16}$O, $^{17}$O and $^{18}$O were measured by Secondary Ion Mass Spectrometry (SIMS) \cite{SIMS_2} at the Department of Physics and Astronomy of the University of Padua (Italy), using a CAMECA IMS-4f spectrometer. A 14.5 keV, 3 nA Cs$^+$ primary beam was rastered over a crater area of 100$\times$100 $\mu$m$^2$ in order to continuously sputter the target surface. During sputtering, the secondary ions $^{16}$O$^-$, $^{17}$O$^-$ and $^{18}$O$^-$ were collected only from a central region of 30 $\mu$m diameter in order to avoid crater-edge effects. 
High mass resolution (M/$\Delta$M $\simeq 5000$) was used to eliminate mass interferences of $^{17}$O$^-$ and $^{18}$O$^-$ with the molecular secondary ions $^{16}$O$^{1}$H$^-$ and $^{17}$O$^1$H$^-$, respectively.  Measurements of Si samples with purity higher than 99.999\% were compared to the natural Si isotopic abundance as a check for possible deviations from linearity of the secondary ions electron multiplier detector.

\begin{figure}[!htb]
\begin{center}
\includegraphics[width=\columnwidth]{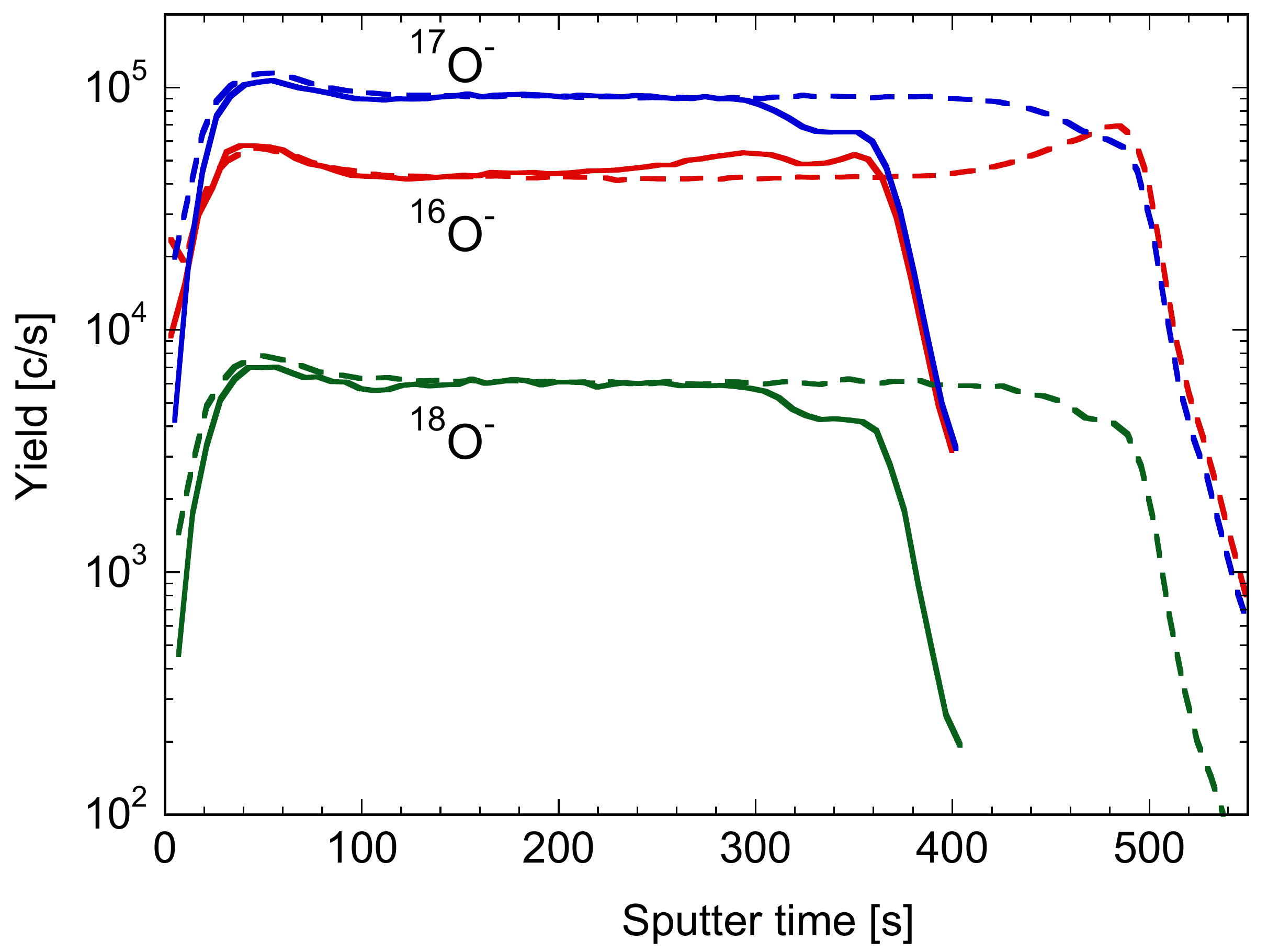}
\end{center}
\caption{(Colour online) Depth yield profiles of secondary $^{16}$O$^-$, $^{17}$O$^-$ and $^{18}$O$^-$ ions as a function of the erosion time, as obtained by Secondary Ion Mass Spectrometry analyses of target 24. Measurements were taken in the region of the target exposed to the proton beam at LUNA (beam-spot, continuos lines) and in peripheral regions not exposed to the proton beam (out-spot, dashed lines). Deterioration of the target thickness is indicated by the drop in (solid line) yields occurring at earlier times with respect to the out-spot measurement. (see text for further details).}
\label{fig:SIMS2}
\end{figure}

In order to assess the extent of possible isotopic abundance variation caused by the proton beam during the measurements at LUNA, comparative SIMS analyses were made in beam-spot and out-spot regions of the targets. 
Fig. \ref{fig:SIMS2} shows the $^{16}$O$^-$, $^{17}$O$^-$ and $^{18}$O$^-$ yield profiles as a function of the erosion time obtained for a target enriched in $^{17}$O (here, target 24). Even though an additional contamination in $^{16}$O was observed in all targets during beam-spot measurements, its effect on the total thickness remained negligible.

Yield curves show a plateau that remains flat throughout the SIMS procedure, indicating that the relevant isotopic content remained uniform in depth and constant in time both in beam-spot and out-spot regions of the targets. The expected target thickness reduction is indicated by the drop in the beam-spot yield occurring at earlier sputter times compared to the out-spot case, in qualitative agreement with results from both RBS and NRRA determinations. 
Anomalous features both at low and high sputtering times can be interpreted as being due either to transients of the sputtering process at the surface of the sample or to the presence of contaminants at the Ta$_2$O$_5$/Ta interface. 
Thus, measurements of isotopic abundance ratios (both in beam-spot and out-spot areas) were taken within the plateau region.
In order to reduce uncertainties, accurate mass calibrations were performed immediately before each isotopic abundance measurement.

Since irregularities in the tantalum backing surface may also affect the reliability of abundances determinations, measurements were performed at several positions (both in beam-spot and out-spot regions) in order to evaluate  local variations of the isotopic concentrations. Average values from such measurements are summarised in Table \ref{tab:SIMS} for a representative sample of targets. The quoted uncertainties represent standard deviations.
For total accumulated charges on target typically below about 25 C the oxygen isotopic concentrations, as measured in the middle of the layer thickness, remain constant within the uncertainties.
For the measurements of the proton-induced reactions at LUNA only targets with total accumulated charges less than about 25 C were used.  

\begin{table}
\centering
\caption{Average values and associated uncertainties (in table header) for isotopic oxygen abundances as measured by SIMS in beam-spot and out-spot regions of a representative sample of targets. Values were taken at middle depths of the Ta$_2$O$_5$ layers within the plateau region (see fig. \ref{fig:SIMS2} and text for further details). }
\begin{tabular}[!htb]{lcccc}
\hline \hline
Target &  $^{16}$O (\%) & $^{17}$O (\%) & $^{18}$O (\%)  & Q$_{\rm tot}$ [C]\\
	& $\pm$0.6 & $\pm$0.5 & $\pm$0.1  & \\
\hline
28 (beam-spot) & 29.7 & 66.1 & 4.2  & 12\\
28 (out-spot) & 30.4 & 65.3 & 4.3  & \\
\hline						
24 (beam-spot) & 30.6 & 65.2 & 4.2 & 23 \\
24 (out-spot) & 29.9 & 65.7 & 4.3  &\\
\hline
29 (beam-spot) & 37.1 & 59.1 & 3.8 & 28\\
29 (out-spot) 	& 31.2 & 64.5 & 4.3  & \\
\hline
30 (beam-spot) & 36.9 & 59.3 & 3.8 & 35\\
30 (out-spot) 	& 31.5 & 64.3 & 4.1 & \\
\hline \hline
\end{tabular}
\label{tab:SIMS}
\end{table}

\section{Conclusions}

Solid targets of known thickness, isotopic composition, and stoichiometry are a key ingredient for nuclear astrophysics studies of proton-induced reactions at sub-Coulomb energies. 

For the study of the $^{17}$O(p,$\gamma$)$^{18}$F and the $^{18}$O(p,$\gamma$)$^{19}$F reactions at the LUNA facility, isotopically enriched Ta$_2$O$_5$ targets have been produced by anodic oxidation of tantalum backings, preliminarily etched in a bath of citric acid.
Targets of various thicknesses (1200-5800 {\rm \AA}) and composition (up to 66\% and 96\% in $^{17}$O and $^{18}$O, respectively) were produced following a standard procedure reported in \cite{Vermy1953,Pringle1972}. Best growth rates and homogeneity were achieved for a temperature $T=25.0 \pm 0.1 ^{\rm o}$C and for a constant current density of 2mA/cm$^2$ in the voltage range 40-220 V ({\em i.e.} growth rate $\sim 32 {\rm \AA/V}$).

Target thickness was monitored during heavy proton-beam bombardment by performing thick-target yield measurements of the well-known $^{18}$O(p,$\gamma$)$^{19}$F resonance at $E_{\rm lab} = 151$ keV. A linear decrease in target thickness as a function of the total accumulated charge was observed for all targets, typically at a rate of 0.2 keV/C. Target thickness degradation amounted to about 20\% after a total charge of 25 C accumulated on target. 

Target stoichiometry and isotopic compositions were carefully investigated by means of RBS and SIMS techniques, respectively. In order to assess any possible variations of these properties during beam bombardment, measurements were carried out both in regions of the targets exposed to the beam and in regions where beam exposure was negligible. RBS results indicated a target stoichiometry Ta/O of $0.39\pm0.02$ in agreement with expectations for targets prepared by anodic oxidation of tantalum backings. The target stoichiometry did not change appreciably during proton beam exposure. The SIMS studies revealed that the relative isotopic abundances remained constant in depth and time throughout the proton-induced reaction measurements carried out at LUNA.

We conclude that the procedure followed to prepare isotopically enriched Ta$_2$O$_5$ targets is very well suited for the production of targets with known and stable composition and stoichiometry. As such, these targets fulfil the key requirements of solid targets for nuclear astrophysics applications.

\section*{Acknowledgments}

The authors are indebted to the INFN mechanical workshop, electronics and chemical laboratories of LNGS. Renzo Storti (University of Padova) is acknowledged for technical assistance. One of the authors (FS) acknowledges financial support from the SUPA Distinguished Visitors Programme.
Financial support by INFN, OTKA (K101328), DFG (BE 4100/2--1), and NAVI is
gratefully acknowledged. A.\,D.\,L. acknowledges financial support by MIUR (FIRB RBFR08549F).  A.\,C.\, acknowledges financial support by Fondazione Cassa di Risparmio di Padova e Rovigo.

%
\bibliographystyle{epjc}

\end{document}